\documentclass[a4paper]{jpconf}
\usepackage{graphicx}
\usepackage[utf8x]{inputenc}
\begin{document}
\title{Latest results from the NA61/SHINE experiment}

\author{Dag Toppe Larsen for the NA61/SHINE Collaboration}

\address{Uniwersytet Jagielloński, Łojasiewicza 11, 30-348 Kraków, Poland}

\ead{dlarsen@cern.ch}

\begin{abstract}
The NA61/SHINE experiment at the CERN Super Proton Synchrotron is pursuing a rich programme on strong interactions, which covers the study of the onset of deconfinement and aims to discover the critical point of strongly interacting matter by performing an energy and system-size scan over the full CERN SPS beam momentum range.
So far the scans of p+p, p+Pb, Be+Be, and Ar+Sc interactions have been completed, samples of Pb+Pb data at three energies have already been taken and Xe+La collisions will be registered this year.

Results from the different reactions are now emerging, in particular the energy dependence of hadron spectra and yields as well as fluctuations.
This contribution presents status and preliminary results from this effort, as well as an outlook for future extensions of the strong interactions programme.
\end{abstract}

\section{Introduction}
NA61/SHINE is a fixed target experiment at the CERN SPS.
Tracking is provided by eight time projection chambers.
Additionally, particle identification is aided by time-of-flight detectors. A modular calorimeter is used to determine the collision centrality for nucleus-nucleus collisions.
Both primary and secondary beams are available to the experiment, allowing data taking with projectile sizes ranging from proton to Pb, as well as with $\pi$ and $K$ mesons.
Besides studying strong interactions, the experiment also performs precise hadron production reference measurements for neutrino (Fermilab and T2K) and cosmic-ray (KASCADE and Pierre Auger Observatory) physics.

The main goal of the strong interactions programme of the NA61/SHINE experiment is to discover the critical point of strongly interacting matter and study the properties of the onset of deconfinement.
The programme is motivated by the discovery of the onset of deconfinement in Pb+Pb collisions at 30$A$~GeV/c by the NA49 experiment.
To achieve this goal a two-dimensional scan of the phase diagram of strongly interacting matter is performed by varying the beam momentum (13$A$-150/158$A$~GeV/c) and the size of the colliding nuclei (p+p, p+Pb, Be+Be, Ar+Sc, Xe+La, Pb+Pb).

\section{Charged $K$ spectra}
\begin{figure}
\includegraphics[width=0.27\linewidth]{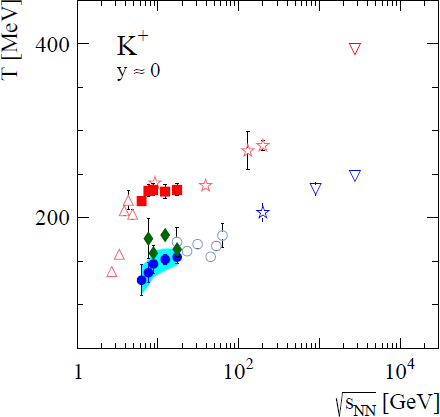}
\hspace{0.01\linewidth}
\includegraphics[width=0.27\linewidth]{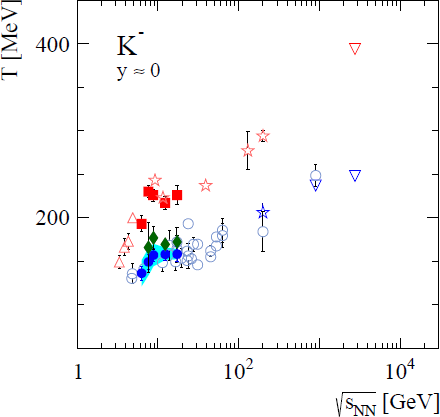}
\hspace{0.01\linewidth}
\includegraphics[width=0.14\linewidth]{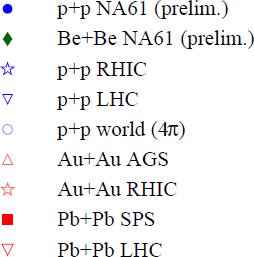}
\begin{minipage}[b]{0.27\linewidth}
\caption{\label{a}Inverse slope parameter $T$ of $m_T$ spectra of $K^+$ (left) and $K^-$ (right) in A+A interactions versus interaction energy.}
\end{minipage}
\end{figure}

NA61/SHINE has recently measured charged $K$ production in violent Be+Be and Ar+Sc collisions at mid-rapidity. 
The Be+Be analysis was performed for the 20\% of events with the smallest forward energy deposited in the PSD detector - the 20\%  most violent collisions.
The tof-dE/dx method of particle identification was used (measurement in a wider rapidity range, via the dE/dx method, is in progress).
Transverse momentum spectra were fitted~\cite{Magda_BNL2017, Tanja_SQM2017} with an exponential function in $m_T$ and the resulting inverse slope parameters $T$ are plotted in Fig.~\ref{a}.
A step followed by a plateau is observed in the energy dependence of the inverse slope parameter in central Pb+Pb collisions, which was predicted as a signature of the onset of deconfinement~\cite{Gazdzicki:1998vd}.
NA61/SHINE results on violent Be+Be collisions and inelastic p+p interactions are also consistent with such a plateau at SPS energies, though at a lower value of $T$.

\begin{figure}
\includegraphics[width=0.27\linewidth]{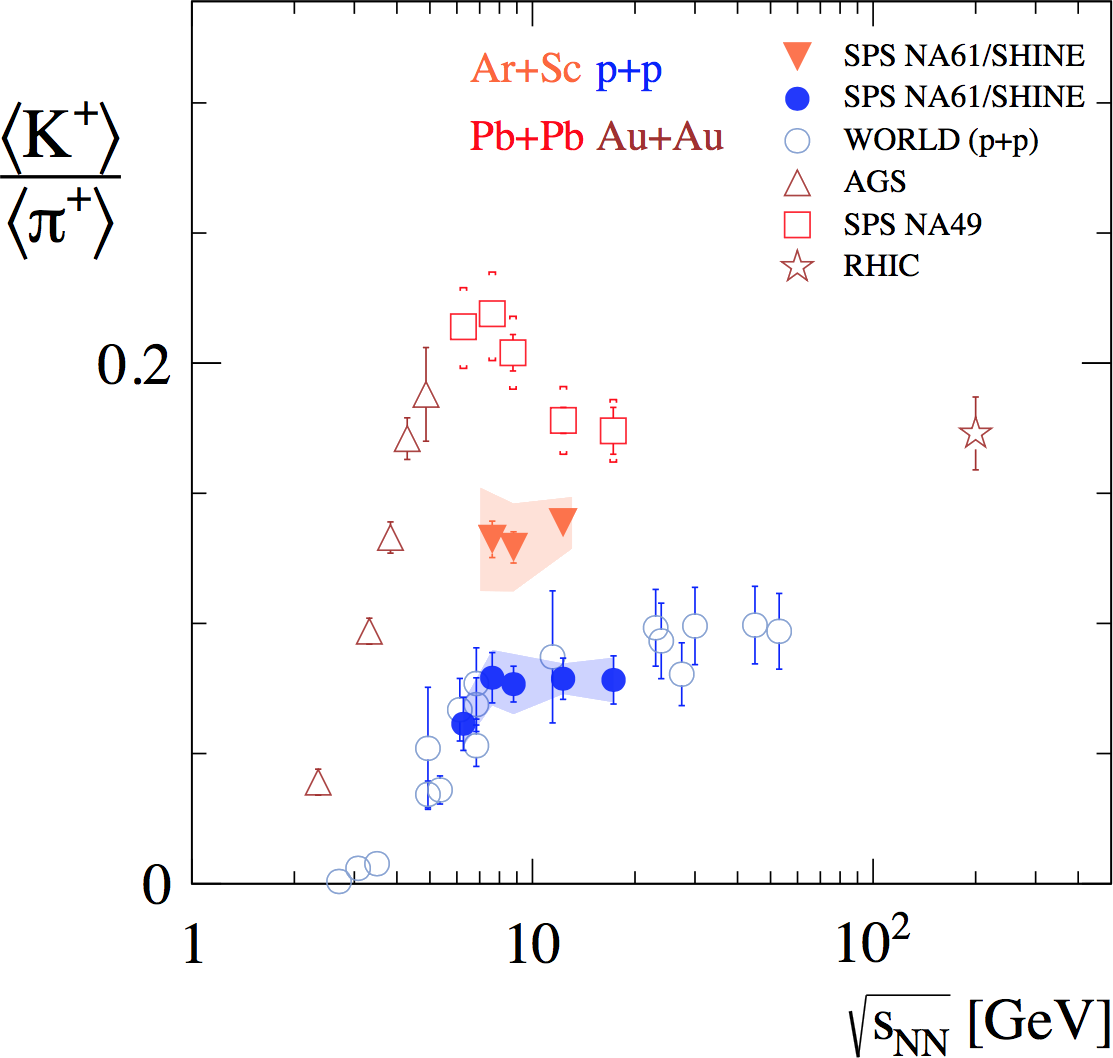}
\hspace{0.01\linewidth}
\includegraphics[width=0.27\linewidth]{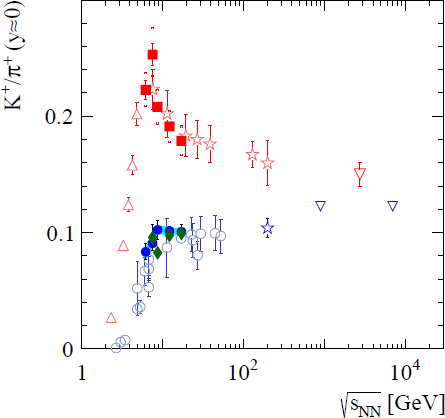}
\hspace{0.01\linewidth}
\includegraphics[width=0.14\linewidth]{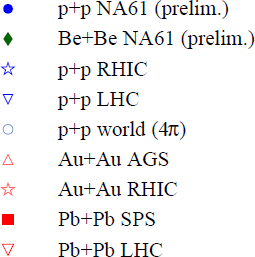}
\begin{minipage}[b]{0.27\linewidth}
\caption{\label{b}Ratios of 4$\pi$ yields ⟨$K^+$⟩/⟨$\pi^+$⟩ (left) and mid-rapidity yields $K^+$/$\pi^+$ (right) in A+A interactions versus interaction energy.}
\end{minipage}
\end{figure}

Figure~\ref{b} (right) shows rapid change in the energy dependence of the $K^+/\pi^+$ ratio observed in Pb+Pb collisions (horn), which was also predicted as a signature of the onset of deconfinement~\cite{Gazdzicki:1998vd}.
In violent Be+Be collisions, as well as in inelastic p+p interactions, a step-like structure with a following plateau is observed in the energy dependence of the $K^+/\pi^+$ ratio, however, at a much lower level.

Charged $K$ spectra and yields for Ar+Sc interactions were measured at beam momenta of 30$A$, 40$A$ and 75$A$~GeV/c.
Double differential $p_T$, y spectra were obtained using the dE/dx method of particle identification.
Then, spectra extrapolated in $p_T$ and rapidity were integrated to derive mean multiplicities (in 4$\pi$) as seen in Fig.~\ref{b} (left).
There is no clear energy dependence and no horn structure visible in Ar+Sc collisions.
The ⟨$K^+$⟩/⟨$\pi^+$⟩ ratio in Ar+Sc collisions is between the ratio in inelastic p+p interactions and the one in violent Pb+Pb collisions.
The analysis of Ar+Sc data at 13$A$, 19$A$ and 150$A$~GeV/c is in progress.

\section{Multiplicity fluctuations}
Multiplicity fluctuations were measured for negatively charged hadrons in Be+Be and Ar+Sc collisions. 
For cross-check, two different measures were used: the scaled variance of the multiplicity distribution, $\omega$[N], and the strongly intensive fluctuation measure $\Omega$[N,E$_P$] calculated for the distributions of multiplicity N and energy of projectile participants E$_P$.
As expected $\omega$[N] for very violent (0-1\% Be+Be) collisions coincides with $\Omega$[N,E$_P$] for violent (0-10\% Be+Be) collisions \cite{Seryakov}.

\begin{figure}
\includegraphics[width=0.32\linewidth]{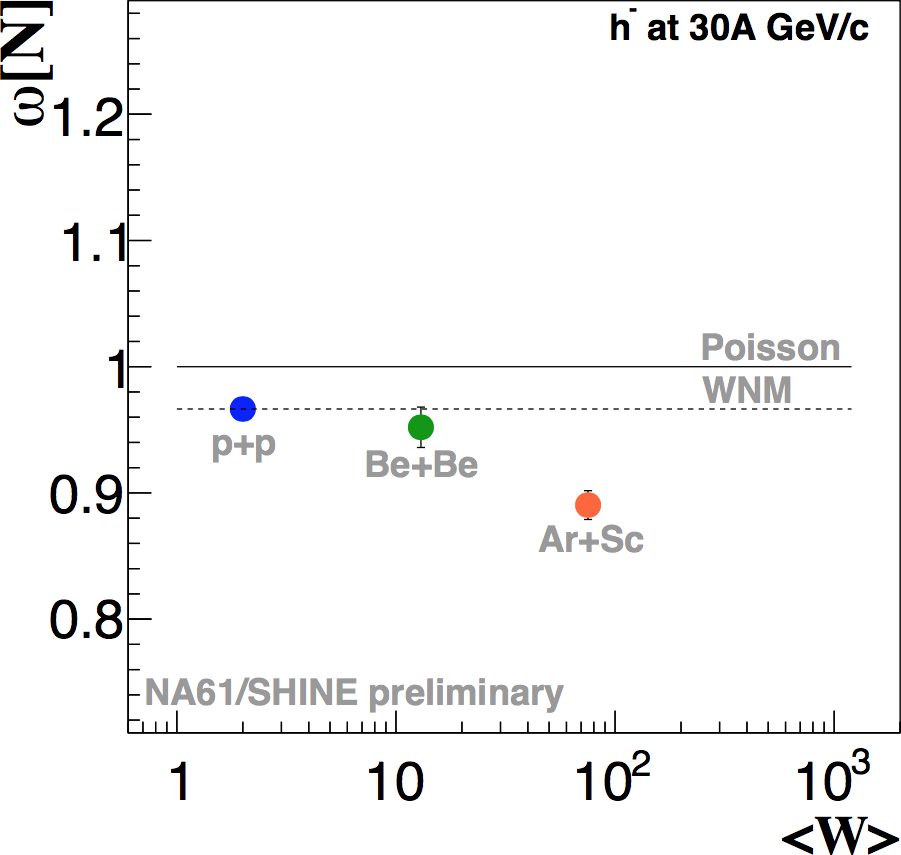}
\hspace{0.01\linewidth}
\includegraphics[width=0.32\linewidth]{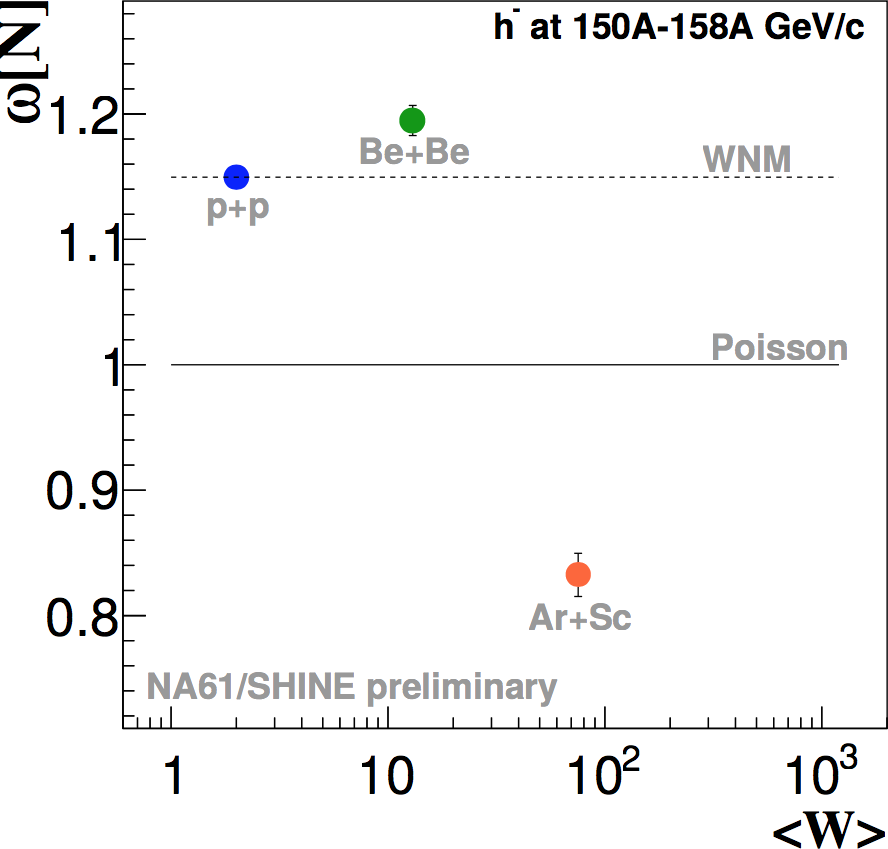}
\hspace{0.01\linewidth}
\begin{minipage}[b]{0.32\linewidth}
\caption{\label{d}Scaled variance $\omega$[N] of the multiplicity distributions of h$^-$ in p+p, Be+Be and Ar+Sc interactions at beam momenta of 30$A$ (left) and 150/158$A$ (right) GeV/c versus the mean number of wounded nucleons $W$.}
\end{minipage}
\end{figure}

Figure~\ref{d} presents the system size dependence of $\omega$[N] for negatively charged hadrons at 30$A$ and 150/158$A$~GeV/c and shows an interesting effect.
A rapid decrease of $\omega$[N] when moving from violent Be+Be to Ar+Sc collisions is observed.
Within the Wounded Nucleon Model $\omega$[N]$_{AA}$=$\omega$[N]$_{pp}$ provided the number of wounded nucleons W does not fluctuate (dotted lines in Fig.~\ref{d}).
The Wounded Nucleon Model with W fluctuations results in $\omega$[N]$_{AA}$$>$$\omega$[N]$_{pp}$.
Thus Ar+Sc results are in qualitative disagreement with predictions of the Wounded Nucleon Model.
For an Ideal Boltzmann gas in the Grand Canonical Ensemble (IB-GCE) formalism $\omega$[N]=1 (Poisson multiplicity distribution), independently of the (fixed) system volume.
Volume fluctuations  increase multiplicity fluctuations resulting in $\omega$[N]$>$1. Adding resonance decays and Bose-Einstein statistics would further increase  $\omega$[N].
Consequently $\omega$[N]$_{AA}$$<$1 is forbidden in the IB-GCE and its extensions. The observed small values of $\omega$[N] in violent Ar+Sc collisions may be due to conservation laws acting in a large volume system~\cite{Begun:2006uu}.
In statistical mechanics they are introduced by using canonical and microcanonical ensembles.
Finally, $\omega$[N]≫1, as seen in p+p reactions at 158$A$ GeV/c, can be understood in statistical models as a result of fluctuations of the fireball volume and/or the energy converted into particle production~\cite{Begun:2008fm}.

\section{System size dependence of particle yields}
\begin{figure}
\includegraphics[width=0.32\linewidth]{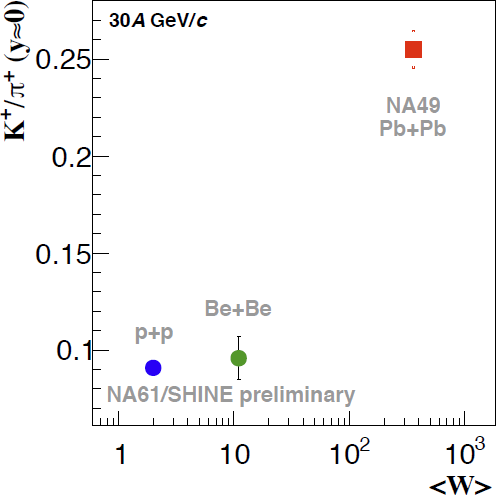}
\hspace{0.01\linewidth}
\includegraphics[width=0.32\linewidth]{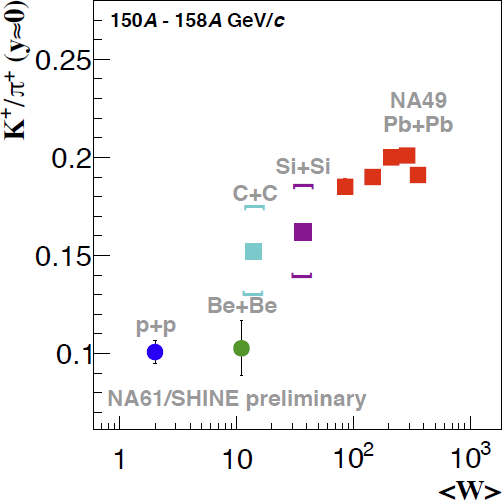}
\hspace{0.01\linewidth}
\begin{minipage}[b]{0.32\linewidth}
\caption{\label{e}Yield ratio K$^+/\pi^+$ at mid-rapidity in p+p, Be+Be and Ar+Sc interactions at beam momenta of 30$A$ (left) and 150/158$A$ (right) GeV/c versus mean number of wounded nucleons.}
\end{minipage}
\end{figure}

Figure~\ref{e} shows example plots of the system size dependence of the ratio of $K^+$ and $\pi^+$ yields at mid-rapidity at two SPS energies.
The scaled variance of multiplicity distributions were already shown in Fig.~\ref{d}.
The Be+Be results are very close to those from p+p interactions independently of collision energy.
Moreover, the data show a jump between light (p+p, Be+Be) and intermediate/heavy (Ar+Sc, Pb+Pb) systems.
The $K^+$/$\pi^+$ ratio in p+p interactions is below the predictions of statistical models.
However, the ratio in central Pb+Pb collisions is close to statistical model predictions for large volume systems~\cite{Becattini:2005xt}.
In p+p interactions, and thus also in Be+Be collisions, multiplicity fluctuations are larger than predicted by statistical models (see Fig.~\ref{d}).
However, they are close to statistical model predictions for large volume systems in central Ar+Sc and Pb+Pb collisions~\cite{Begun:2006uu}.
Thus the observed rapid change of hadron production properties that start when moving from Be+Be to Ar+Sc collisions can be interpreted as the beginning of creation of large clusters of strongly interacting matter.

Furthermore hadron production properties in heavy ion collisions were found to change rapidly with increasing collision energy in the low SPS energy domain, $\sqrt{s_{NN}}$≈10 GeV (for a recent review see Ref.~\cite{Gazdzicki:2014sva}).
The results shown in Figures~\ref{a} and~\ref{b} indicate that this is also the case in inelastic p+p interactions and probably also in Be+Be collisions.
The phenomenon in collisions of heavy nuclei is labelled as the onset of deconfinement and interpreted as the beginning of creation of quark-gluon plasma with increasing collision energy~\cite{Gazdzicki:2010iv}.

\section{$\phi$ meson production in p+p collisions}

\begin{figure}
\includegraphics[width=0.31\linewidth]{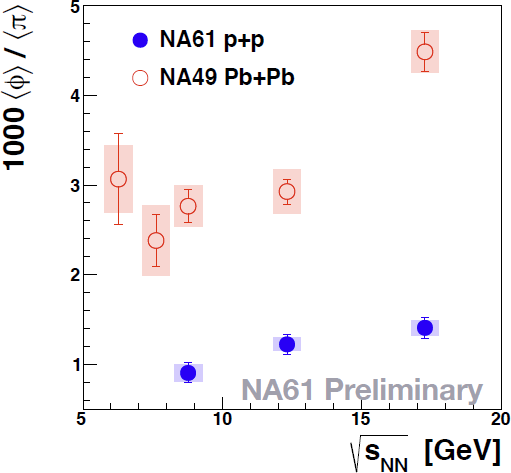}
\hspace{0.01\linewidth}
\includegraphics[width=0.33\linewidth]{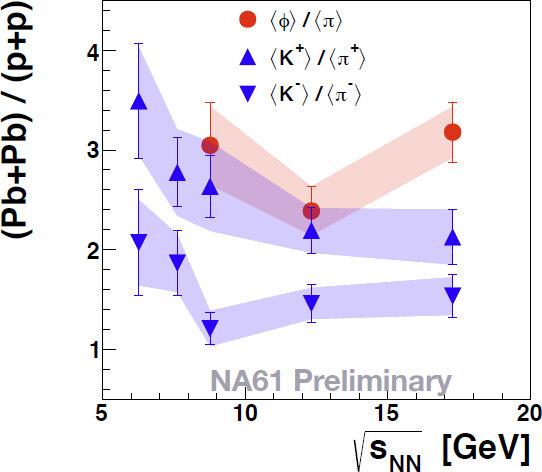}
\hspace{0.01\linewidth}
\begin{minipage}[b]{0.32\linewidth}
\caption{\label{f}Yield ratio ⟨$\phi$⟩/⟨$\pi$⟩ in p+p and central Pb+Pb interactions (left) and double ratio (Pb+Pb)/(p+p) of ⟨$\phi$⟩/⟨$\pi$⟩ and ⟨$K^±$⟩/⟨$\pi^±$⟩ (right) versus centre of mass energy per nucleon pair $\sqrt{s_{NN}}$.}
\end{minipage}
\end{figure}

New results on $\phi$ meson production were obtained in inelastic p+p interactions at 40, 80, and 158$A$~GeV/c. 
The ratio of mean multiplicity of $\phi$ and $\pi$ mesons in p+p and central Pb+Pb collisions as a function of centre of mass energy per nucleon pair $\sqrt{s_{NN}}$ is shown in Figure~\ref{f} (left), while the double ratio (Pb+Pb)/(p+p) of the ratios ⟨$\phi$⟩/⟨$\pi$⟩ and ⟨$K^±$⟩/⟨$\pi^±$⟩ is presented in Fig.~\ref{f} (right).
As seen from Fig.~\ref{f} (left) the ⟨$\phi$⟩/⟨$\pi$⟩ ratio increases with $\sqrt{s_{NN}}$.
In Pb+Pb collisions it is approximately three times larger than in p+p collisions, independently of the interaction energy.
Figure~\ref{f} (right) shows that the enhancement between p+p and Pb+Pb of the double ratio ⟨$\phi$⟩/⟨$\pi$⟩ is close to that of ⟨$K^+$⟩/⟨$\pi^+$⟩ and systematically larger than that of ⟨$K^-$⟩/⟨$\pi^-$⟩.
More information can be found in Ref.~\cite{Tanja_SQM2017}.

\section{Open Charm measurements}
The strong interactions programme of the NA61/SHINE experiment at the CERN SPS has been expanded to allow precise measurements of particles with short lifetime.
The study of open charm meson production provides an important tool for new detailed investigations of the properties of hot and dense matter formed in nucleus-nucleus collisions.
In particular, it opens new possibilities for studies of such phenomena as in-medium parton energy loss and quarkonium dissociation and possible regeneration, thus bringing new information to probe deconfinement.

\begin{figure}
\includegraphics[width=0.32\linewidth]{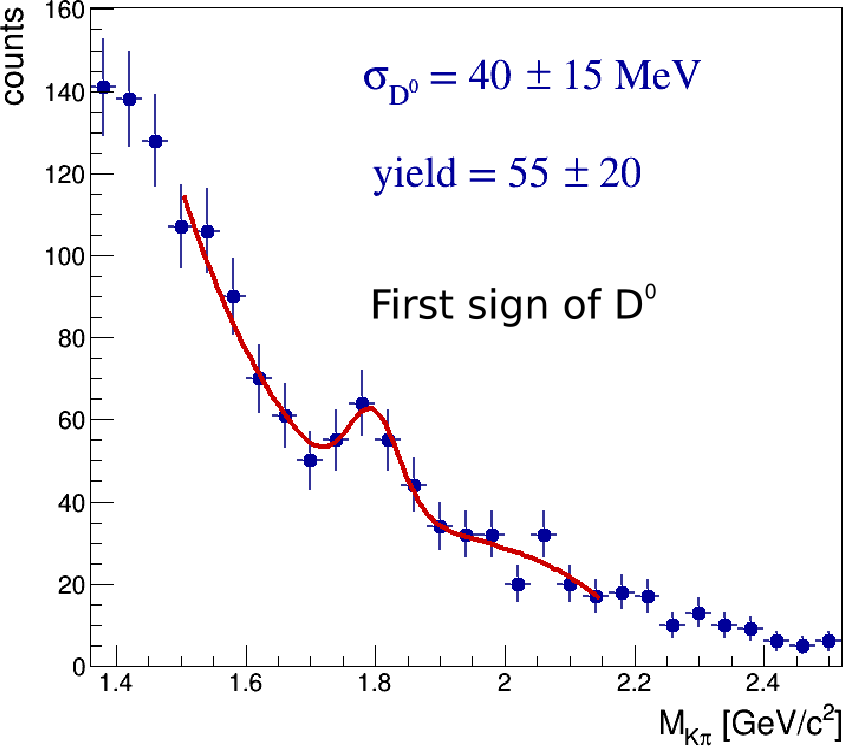}
\hspace{0.01\linewidth}
\begin{minipage}[b]{0.32\linewidth}
\caption{\label{g}First indication of D$^0$ production in the $+\pi$ decay channel in central Pb+Pb interactions at beam momentum of 150$A$ GeV/c.}
\end{minipage}
\end{figure}

A new vertex detector was constructed for the NA61/SHINE experiment for measurements of the very rare processes of open charm production in nucleus-nucleus collisions at the SPS. It was designed to meet the challenges of precise track registration and of high spatial resolution in primary and secondary vertex reconstruction.
A small-acceptance version of the vertex detector, SAVD (Small Acceptance Vertex Detector), was installed last year with a Pb target in the Pb beam of 150$A$ GeV/c momentum, and a modest set of data were collected.
The main goal of the ongoing data analysis was to observe a signal from D$^0$ mesons.
Figure~\ref{g} shows the first indication of D$^0$ production in the $K+\pi$ decay channel~\cite{Anastasia}.

\section*{Acknowledgements}
This analysis was partially supported by the Polish National Science Centre, grants 2014/15/B/ST2/02537 and 2015/18/M/ST2/00125.

\section*{References}
\bibliography{na61References}
\bibliographystyle{na61Utphys}

\end{document}